\documentclass[twocolumn,showpacs,preprintnumbers,amsmath,amssymb]{revtex4}

\usepackage{graphicx}
\usepackage{dcolumn}
\usepackage{bm}

\begin{document}


\title{Spatiotemporal Model for Kerr Comb Generation \\
      in Whispering Gallery Mode Resonators}

\author{Yanne K. Chembo}
\thanks{E-mail: yanne.chembo@femto-st.fr}
\affiliation{FEMTO-ST Institute [CNRS UMR6174], Optics Department, \\
             16 Route de Gray, 25030 Besan\c con cedex, FRANCE.}
\author{Curtis R. Menyuk}
\affiliation{University of Maryland, Baltimore County, Department of Computer Science and Electrical Engineering, \\
             1000 Hilltop Circle, Baltimore, Maryland 21250, USA}
\date{\today}

\begin{abstract}

We establish an exact partial differential equation to model Kerr comb generation in whispering-gallery mode resonators.
This equation is a variant of the Lugiato-Lefever equation that includes higher-order dispersion and
nonlinearity. This spatio-temporal model, whose main variable is the total intracavity field,
is significantly more suitable than the modal expansion approach for the theoretical understanding and the numerical simulation of wide-span combs. 
It allows us to explore pulse formation in which a large number of modes interact cooperatively.
This versatile approach can be straightforwardly extended to include higher-order dispersion, as well as other phenomena like Raman, Brillouin and Rayleigh scattering. We demonstrate for the first time that when the dispersion is anomalous, Kerr comb generation can arise as the spectral signature of dissipative cavity solitons, leading to wide-span combs with low pumping.
\end{abstract}

\maketitle

The development of frequency combs -- equidistant frequency lines
 from a short-pulse laser -- revolutionized the measurement of
 frequencies~\cite{Cundiff_Ye} and has opened up a host of potential applications in
 fundamental and applied physics, including the measurement of physical
 constants, the detection of earth-like planets, chemical sensing, the
 generation, measurement, and distribution of highly accurate time, and
 the generation of low-phase-noise microwave radiation~\cite{Diddams}.
 Ti:sapphire lasers were used as the original source of frequency
 combs, but in the past seven years, alternative fiber laser sources
 have developed~\cite{Diddams}.  Recently, Del'Haye, et al.~\cite{DelhayeKipp} demonstrated that
 it is possible to use the whispering gallery modes in microresonators
 in combination with the Kerr effect to generate an equidistant
 frequency comb that is also referred to as a Kerr comb.  Since many
 applications in both fundamental and applied science would benefit
 from the small size, simplicity, robustness, and low power consumption
 of these whispering gallery mode sources, a considerable worldwide effort
 has gone into understanding and controlling them~\cite{Review_Kerr_combs_Science}.  In particular,
 there have been several efforts to develop mathematical models of
 these sources~\cite{YanneNanPRL,YanneNanPRA,GaetaOE,Matsko}, but all the efforts to date have serious drawbacks.

 A complete modal expansion has been derived to describe the growth of
 the Kerr combs from noise~\cite{YanneNanPRL,YanneNanPRA}.  This model predicts a cascaded
 growth in which a primary comb is first generated, which then generates
 a secondary comb, and later higher order combs.  This model is in complete agreement with
 experiments~\cite{YanneNanPRL}.  However, it is difficult and computationally
 expensive to use this mode expansion beyond the primary comb
 generation because the number of modes that must be kept in a
 calculation grows like the third power of time.  Moreover, it is
 difficult to study pulse formation in this model, since a
 large number of modes interact cooperatively.  Pulse formation plays a
 critical role in comb generation.  It is desirable to find a
 spatio-temporal model that would be analogous to the Haus modelocking
 equation (a variant of the nonlinear Schr\"odinger equation) that has
 been used with great success to study modelocked lasers~\cite{Haus}.
 However, this equation is not itself appropriate for whispering gallery mode
 resonators.  This equation assumes that pulses are isolated, and it does
 not take into account the periodic boundary conditions and the
 fundamental role that continuous wave pumping and dissipation play in
 the evolution~\cite{Matsko}.
 
 In this letter, we show that the Lugiato-Lefever equation (LLE)~\cite{LL}
 is the appropriate spatio-temporal model of whispering gallery mode
 resonators.  We demonstrate that the previously-derived modal
 expansion is equivalent to a variant of the LLE that includes
 higher-order dispersion and nonlinearity, and the standard LLE closely
 approximates the modal expansion.  We then use this model to study
 pulse formation and determine conditions under which multiple pulses
 (rolls) and single pulses (dissipative solitons) form in the cavity.
 We show that a broadband comb is obtained when a single dissipative
 soliton forms.

 The LLE can be considered a variant of the nonlinear Schr\"odinger
 equation (NLSE) that includes damping, driving, and detuning.  It has
 been the subject of intensive mathematical study~\cite{Scroggie,Japanese_Group,Kozyreff}, and it has
 been experimentally demonstrated that this model can be applied to
 fiber ring resonators~\cite{Temporal_CS}.  Unexpected phenomena such has convective instability~\cite{IFISC1} and excitability~\cite{IFISC2}
 are also known occur in this system. 
 The identification of the LLE as the fundamental equation governing the field evolution in
 whispering gallery mode resonators allows us to take advantage of this
 prior work, while at the same time suggesting efficient computational
 methods for studying the field evolution, including Kerr comb generation.

Our starting point is the modal equations that were derived in ref.~\cite{YanneNanPRA}.
The modal structure of the whispering gallery mode resonators is
well-understood.  It has been shown that these resonators can sustain
several families of eigenmodes that trap light inside the cavity by
total internal reflection.  We are only interested in the fundamental
family of modes that can be characterized by their toroidal structure.
Provided that the polarization is fixed, the members of this family are
unambiguously defined by an integer wavenumber $\ell$ that characterizes
each member's angular momentum and can be interpreted as the total
number of reflections that a photon makes during one round trip in the
cavity.

We denote the eigennumber of the pumped mode as $\ell_0$.  If we restrict
ourselves to the spectral neighborhood of $\ell_0$, the eigenfrequencies
can be expanded in a Taylor series, whose first three elements are
\begin{eqnarray}
\omega_\ell = \omega_{\ell_0} + \zeta_1  (\ell - \ell_0) + \frac{1}{2} \zeta_2  (\ell - \ell_0)^2   \, ,
\label{frequency_expansion}
\end{eqnarray}
where
 $ \omega_{\ell_0} \equiv \omega_0$ is the eigenfrequency at $\ell = \ell_0$, $\zeta_1
=\left. d\omega/d \ell\right|_{\ell=\ell_0} = \Delta \omega_{\rm FSR}$ is the
free-spectral range of the resonator or the intermodal angular
frequency, and $\zeta_2 = \left. d^2\omega/d \ell^2\right|_{\ell=\ell_0}$
is the second-order dispersion coefficient.  The quantity $\zeta_2$
corresponds to $\zeta$ in refs.~\cite{YanneNanPRA,YanneNanPRL}.  In the case of a disk
resonator with main radius $a$, we find $\zeta_1 = c/n_0 a$, where $c$
is the velocity of light and $n_0$ is the index of refraction at
$\omega_0$.  This intermodal angular frequency is linked to the
round-trip period of a photon through the resonator as $T = 2\pi/\zeta_1$.
The second-order dispersion $\zeta_2$ denotes the lowest-order
deviation from frequency equistance of the modes.  When $\zeta_2=0$,
the eigenfrequencies are equidistant to lowest order and are separated
by $\zeta_1$. The dispersion is normal when $\zeta_2
<0$ and anomalous when $\zeta_2 > 0$. In fact, $\zeta_2$ is the sum of two contributions
-- the geometrical dispersion (generally normal)
and the material dispersion (which can be normal or anomalous). Explicit expressions are given in refs.~\cite{YanneNanPRA,YanneNanPRL} for a spherical resonator.

We now consider the elements of the modal expansion in a range of $\ell$-values in which the expansion of Eq.~(\ref{frequency_expansion}) is valid.  The slowly varying envelopes ${\cal A}_{\ell}$ obey the equations~\cite{YanneNanPRA,YanneNanPRL}
\begin{eqnarray}
\frac{d {\cal A}_\ell}{dt}  &=& -\frac{1}{2} \Delta \omega_\ell \, {\cal A}_\ell  + \frac{1}{2}  \Delta \omega_\ell \, {\cal F}_{\ell} \, e^{i(\Omega_0 - \omega_\ell ) t} \delta (\ell- \ell_0)  \\
&& -ig_0 \sum_{\ell_m,\ell_n,\ell_p}  {\cal A}_{\ell_m} {\cal A}_{\ell_n}^* {\cal A}_{\ell_p} e^{[i(\omega_{\ell_m} - \omega_{\ell_n} + \omega_{\ell_p} -\omega_{\ell} )t ] } \nonumber \\
&&  \,\,\,\,\,\,\,\,\, \,\,\,\,\,\,\,\,\,\,\,\,\,\,\,\,\,\,\,\,\,\,\, \times \Lambda_{\ell}^{\ell_m \ell_n \ell_p} \delta (\ell_m - \ell_n + \ell_p - \ell) \nonumber \, ,
\label{modal_equations}
\end{eqnarray}
where $\delta (x)$ is the Kr\"onecker delta-function that equals $1$ when
$x=0$ and equals zero otherwise.  The mode fields have been normalized
so that $|A_\ell|^2$ corresponds to the photon number in the mode $\ell$. The
mode bandwidth $\Delta \omega_\ell = \omega_\ell/Q_0$ is inversely
proportional to the loaded quality factor $Q_0$ and to the phonon
lifetime $\tau_{\rm ph, \ell} = 1/\Delta \omega_\ell $. The four-wave mixing gain is
$g_0 = n_2 c\hbar\omega_0^2/n_0^2 V_0$, where $\hbar$ is Planck's
constant, $n_2$ is the Kerr coefficient, and $V_0$ is the effective
mode volume. The coefficient $\Lambda_{\ell}^{\ell_m \ell_n \ell_p}$ indicates the mode
overlap.  The parameter ${\cal F}_0$ denotes the amplitude of the external
excitation, while $\Omega_0$ is the angular frequency of the pump
laser, and it is assumed to be close to $\omega_0$. Typically, resonant
pumping only occurs when $|\Omega_0 - \omega_0| \lesssim \Delta \omega_0$.

The spatiotemporal slowly varying envelope of the total field ${\cal A}(\theta,t)$ may now be written
\begin{eqnarray}
{\cal A}(\theta ,t)  =  \sum_{\ell} {\cal A}_\ell (t)  \exp [i (\omega_\ell - \omega_0)t - i (\ell - \ell_0) \theta ] \, ,
\label{field_expansion}
\end{eqnarray}
where $\theta\in[-\pi,\pi]$ is the azimuthal angle along the
circumference.  From Eq.~(\ref{field_expansion}), it follows that
\begin{eqnarray}
\frac{\partial {\cal A}}{\partial t}  =     \sum_{\ell} \left[  \frac{d {\cal A}_\ell}{d t} + i (\omega_\ell - \omega_0){\cal A}_\ell \right] 
                                                        e^{[i (\omega_\ell - \omega_0)t - i (\ell - \ell_0) \theta ]} \, .
\label{derivative_1}
\end{eqnarray}
Equation~(\ref{field_expansion}) independently yields 
\begin{eqnarray}
 i^n  \frac{\partial^n {\cal A}}{\partial \theta^n} = \sum_{\ell} (\ell -\ell_0)^n {\cal A}_\ell  e^{[i (\omega_\ell - \omega_0)t - i (\ell - \ell_0) \theta ]} \, ,
\label{derivative_1bis}
\end{eqnarray}
so that if we restrict ourselves in the degenerate case where $\Lambda_{\ell}^{\ell_m \ell_n \ell_p} =1$  and $ \Delta \omega_\ell = \Delta \omega_0  $, 
the evolution equation~(\ref{derivative_1}) can be rewritten as 
 \begin{eqnarray}
\frac{\partial {\cal A}}{\partial t} & =&   -\frac{1}{2} \Delta \omega_0 {\cal A} -i g_0 |{\cal A}|^2 {\cal A} + \frac{1}{2} \Delta \omega_0 {\cal F}_{0}  e^{i\sigma t} \nonumber \\
                                      && - \zeta_1 \frac{\partial {\cal A}}{\partial \theta } - i \frac{\zeta_2}{2} \frac{\partial^2 {\cal A}}{ \partial \theta^2 } \, ,
\label{derivative_2}
\end{eqnarray}
where $\sigma = \Omega_0 - \omega_0$ is the detuning between the laser and cavity resonance frequencies.
It is useful to translate the frequency of the carrier
envelope to remove the explicit time-dependence of the driving term by
making the transformation ${\cal A} \rightarrow {\cal A}\exp(i\sigma t)$.  It is
also useful to transform the $\theta$-coordinate to remove the group
velocity motion by making the transformation $\theta \rightarrow \theta
- \zeta_1 t\, {\rm mod}\, [2\pi]$. When that is done, we find that Eq.~(\ref{derivative_2}) becomes
\begin{eqnarray}
\frac{\partial {\cal A}}{\partial t}
&=& - \frac{1}{2} \Delta \omega_0 {\cal A} - i \sigma {\cal A}+ \frac{1}{2} \Delta \omega_0 {\cal F}_0 \nonumber \\
&&  -i g_0 |{\cal A}|^2 {\cal A} - i \frac{\zeta_2}{2} \frac{\partial^2 {\cal A}}{ \partial \theta^2 }  \, .
\label{derivative_3}
\end{eqnarray}
Finally, this equation can be rewritten in the form of the normalized Lugiato-Lefever equation 
\begin{eqnarray}
\frac{\partial \psi}{\partial \tau}  =  - (1 +i\alpha) \psi + i |\psi|^2 \psi - i \frac{\beta}{2} \frac{\partial^2 \psi}{ \partial \theta^2 } +F \, ,
\label{final_eq_dimensionless}
\end{eqnarray}
where the field envelope has been rescaled so that $\psi =
(2g_0/ \Delta \omega_0)^{1/2} {\cal A}^*$ and the time has been rescaled so that
$\tau =  \Delta \omega_0 t/2$.  The dimensionless parameters of this normalized
equation are the frequency detuning $\alpha = -2\sigma/\Delta \omega_0$, the
dispersion $\beta = -2\zeta_2/\Delta \omega_0$, and the external pump $F = (2g_0/\Delta \omega_0)^{1/2}
 {\cal F}_0^*$.  
This LLE with {\it periodic} boundary conditions is the {\it exact} counterpart of the modal expansion as long as higher-order dispersion and the variation of $\Delta\omega_\ell$ and $\Lambda_\ell^{\ell_m \ell_n \ell_p}$ can be neglected.  Since the higher-order corrections can be calculated, it is always possible to check the validity of the LLE, and it will remain valid until the comb is nearly octave-spanning. 
As a consequence, all the conclusions that were previously
obtained using the modal expansion and confirmed experimentally
can also be obtained by solving the LLE.  These two twin-models
are however useful in different and complementary ways.  On the one
hand, the modal expansion must be used to determine threshold phenomena
when a small number of modes are involved.  On the other hand, the LLE
is appropriate to use when many hundreds or thousands of modes interact
since it does not refer to the individual modes.  In particular, as we will show later,
it is useful in the study of Kerr combs or pulse (soliton) growth and
propagation in which a large number of modes interact cooperatively.
\begin{figure}
\begin{center}
\includegraphics[width=8cm]{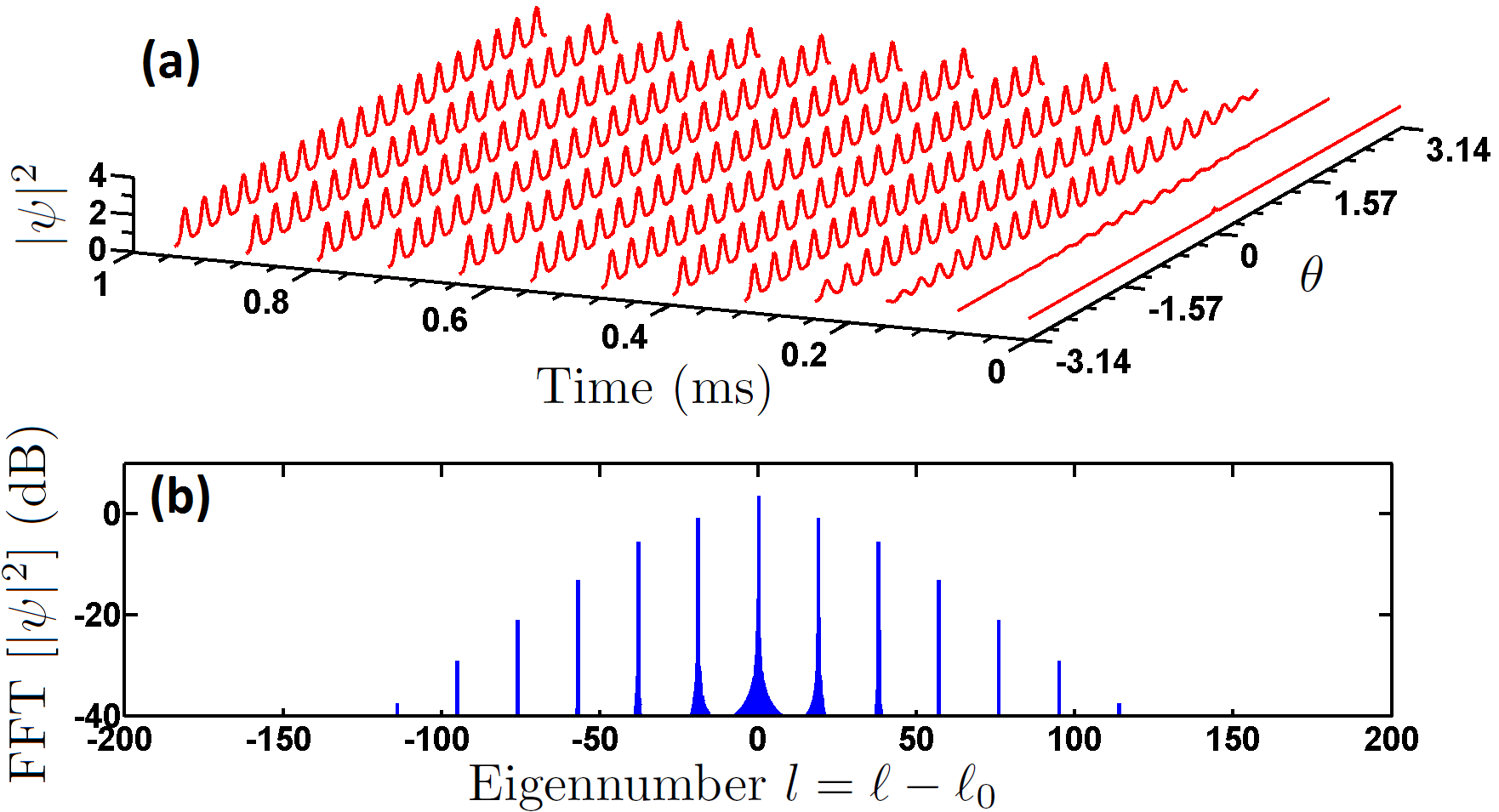}
\end{center}
\caption[Figure2]
{\label{Figure2} Multiple-FSR comb with $\alpha = 0$, $\beta=-0.0125$ and $F=1.71$.  The intracavity pump is $|\psi_0|^2 = 1.2$, so that it is above threshold $|\psi|^2_{\rm th} = 1$. The parameters correspond to Fig.~10b in ref.~\cite{YanneNanPRA}. 
$(a)$ Time-domain dynamics that consists of 19 smooth pulses (``roll'' Turing pattern solution).  
$(b)$ FFT of the roll solution, corresponding to an order 19 multiple-FSR comb.}
\end{figure}

We now discuss the meaning and order of magnitude of the variables and
parameters that appear in the dimensionless LLE.  The detailed
analysis of ref.~\cite{YanneNanPRA} showed that the threshold value for the Kerr comb is
given by $| {\cal A}|_{\rm th}^2 = \Delta \omega_0 /2 g_0$, which exactly corresponds to
$|\psi|_{\rm th}^2 = 1$.  Hence, we would expect $\psi$ to be on the order of 1
in our theoretical analysis and numerical simulations. 
Note that the stationary intra-cavity field $\psi_0$ before comb generation can be obtained by setting all the derivatives to zero 
in Eq.~(\ref{final_eq_dimensionless}).
The dimensionless time can be rewritten as $t/2\tau_{\rm ph}$, where we
recall that $\tau_{\rm ph}$ is the photon lifetime, which is typically
a few $\mu$s in ultra-high-$Q$ whispering gallery mode resonators.
Using the external field's phase as our reference, the amplitude $F$ is
real and positive and $F^2$ is proportional to the external pump power.
The real parameter $\alpha$ equals the ratio of the detuning to half
the linewidth, so that we expect to be off resonance when $|\alpha| >
1$.  This parameter is easily tuned in experiments.  The parameter
$\beta$ equals the ratio of the walkoff due to second-order dispersion
to the half-linewidth.  This parameter is negative for anomalous
dispersion and positive for normal dispersion.  It is a relatively
small parameter.  In refs.~\cite{YanneNanPRA,YanneNanPRL} for example, it was found that
$\beta \sim -0.01$ was sufficient to generate combs.  In practice, the
dispersion magnitude $|\beta|$ should not be too large ($\sim 1$), as it
would mean that the deviation from equidistance of the mode frequencies
would be too strong for the nonlinearity to compensate, and the
emergence of a wide-span Kerr comb would be suppressed.

\begin{figure}
\begin{center}
\includegraphics[width=8cm]{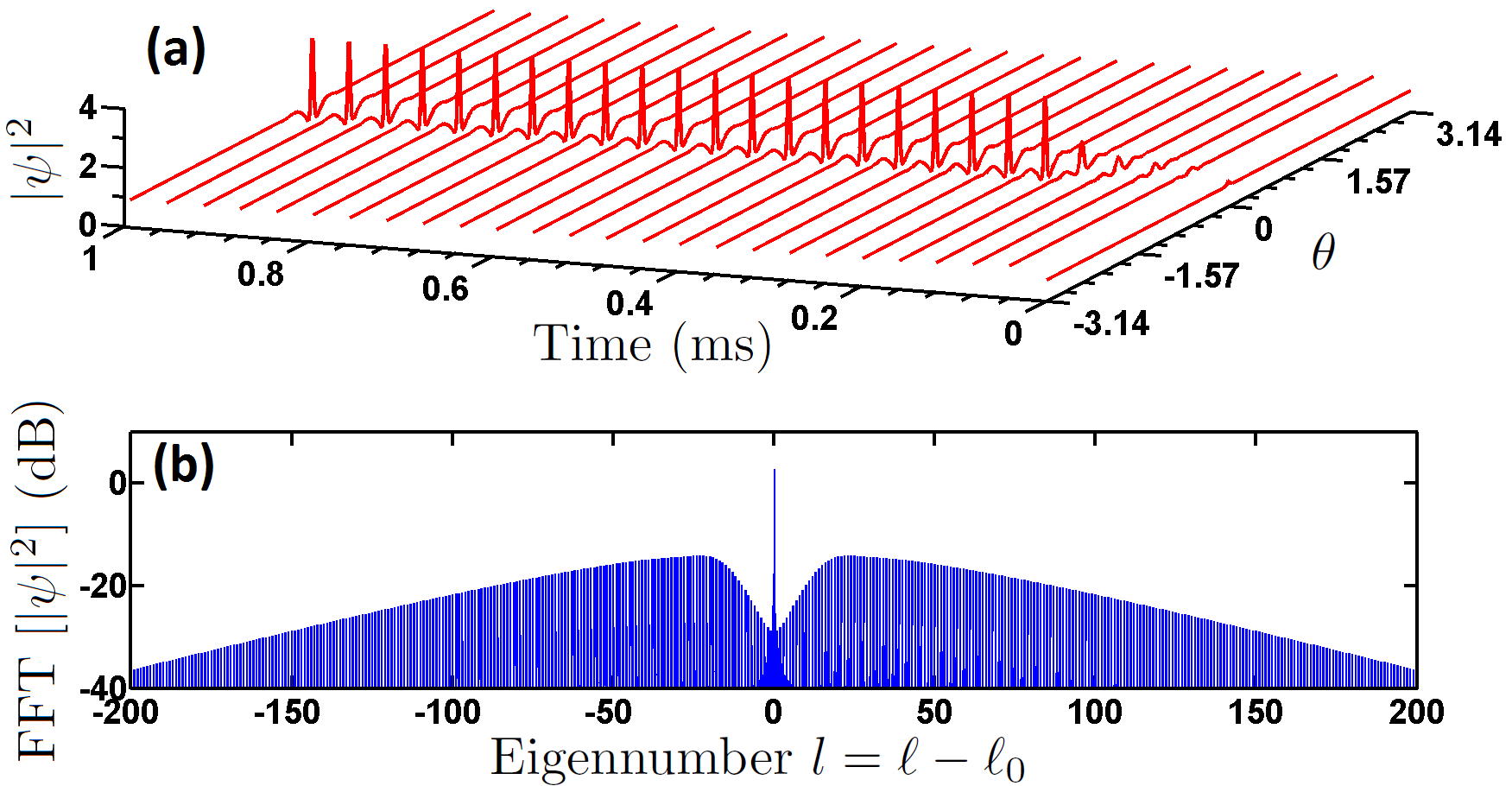}
\end{center}
\caption[Figure3]
{\label{Figure3} Temporal cavity soliton with $\alpha=1.7$, $\beta=-0.002$ and $F=1.22$. 
The intracavity pump is $|\psi_0|^2 = 0.95$, so that it is {\em below} threshold.
$(a)$ Cavity soliton formation. 
$(b)$ FFT of the soliton, corresponding to a wide-span, low-threshold Kerr frequency comb.}
\end{figure}

Numerical simulations were performed using the split-step Fourier
algorithm, which is commonly used in simulations of the 1-D generalized
NLSE.  We note that this simulation method inherently assumes
periodic boundary conditions because it is based on the fast Fourier
transform (FFT).  For studies of pulses in optical fibers -- in which the
pulse duration can often be many orders of magnitude smaller than the
separation between pulses -- this feature is often an annoyance that
leads to boundary-induced artifacts.  In our case, the physical
boundary conditions are periodic.  The split-step Fourier algorithm is
therefore a remarkably cost-effective computational tool, enabling the
simulation of Kerr comb dynamics with a laptop computer in a few
minutes regardless of its spectral span, as opposed to a few days with
the modal expansion for wide-span combs.

For the numerical simulation of the LLE, Eq.~(\ref{final_eq_dimensionless}), we have considered a
calcium fluoride resonator with a radius $a=2.5$ mm.  The polar
eigennumber of the TE-polarized pump mode is $\ell_0 = 14350$,
corresponding to a wavelength of $1560.5$~nm and $\omega_0 =
2\pi \times 192.24$~THz in vacuum.  The index of refraction is $n_0 =
1.43$.  The resonator is critically coupled with a loaded quality
factor $Q_0 = 3\times10^9$, corresponding to a central mode bandwidth
of $\Delta \omega_0 =  \omega_0 /Q \simeq  2\pi \times 64 $~kHz.  The free spectral
range is equal to $\Delta\omega_{\rm FSR} = \zeta_1 = 2\pi\times13.36$~GHz. 

In Fig.~\ref{Figure2}, we present simulation results that correspond to those that
have already been obtained from the modal expansion, namely the multiple-FSR solution.
It can be seen that the solution corresponds to the formation of ``rolls'' in the time domain. 
We take advantage of the LLE formulation to show particular solutions that
would be difficult to observe in a strictly modal study.  Figure~\ref{Figure3}a
shows the formation of a single-peaked cavity soliton.  This
dissipative localized structure is a sub-critical Turing pattern.
The figure corresponds to a single pulse of width $\Delta T\simeq 500$~fs, circulating inside the cavity with a round-trip time $T =
2\pi/\Delta\omega_{\rm FSR} \simeq75$~ps.  In the frequency domain,
this pulse consists of several tens of modelocked whispering gallery
modes, whose frequencies have been nonlinearly shifted so that they are
equidistant, as shown in Fig.~\ref{Figure3}b.  Because this cavity soliton is
subcritical, it emerges abruptly and at a low pump power.  The
solitonic Kerr combs do not grow as the pump power increases; instead,
they are destroyed.  By contrast, super-critical combs like the one shown in Fig.~\ref{Figure2} have spectral
components that grow in number and power as the external pump power
grows.  However, the dissipative cavity solitons are robust, and their
full width at half maximum decreases with the dispersion parameter $|\beta|$.

For the simulation of wide-span combs, the LLE can be generalized to take into account higher-order dispersion ($\zeta_n = \left. d^n\omega/d \ell^n\right|_{\ell=\ell_0} \neq 0$, with $n \geq 3$), non-degeneracy of the bandwidths ($\Delta \omega_\ell \neq \Delta \omega_0$), and spatial non-degeneracy of the eigenmodes ($\Lambda_{\ell}^{\ell_m \ell_n \ell_p} \neq 1$).  
Indeed, these higher-order effects have to be taken into consideration for the simulation of octave-spanning combs.  If we account for third-order dispersion and the first-order variation of $\Lambda_{\ell}^{\ell_m \ell_n \ell_p}$, Eq~(\ref{derivative_2}) becomes
\begin{eqnarray}
\frac{\partial {\cal A}}{\partial t} & =&   -\frac{1}{2} \Delta \omega_0 {\cal A} -i g_0 |{\cal A}|^2 {\cal A} + \frac{1}{2} \Delta \omega_0 {\cal F}_{0}  e^{i\sigma t} \\
                                      && - \zeta_1 \frac{\partial {\cal A}}{\partial \theta } 
                                        - i \frac{\zeta_2}{2} \frac{\partial^2 {\cal A}}{ \partial \theta^2 }
                                          +\frac{\zeta_3}{6} \frac{\partial^3 {\cal A}}{ \partial \theta^3 } \nonumber  \\
                                      && +g_0 \left[ \eta_{\ell} \frac{\partial}{ \partial \theta } ( | {\cal A}|^2 {\cal A})
                                       + 2 \eta_{\ell_m} | {\cal A}|^2  \frac{\partial {\cal A}}{ \partial \theta } 
                                        - \eta_{\ell_n}    {\cal A}^2  \frac{\partial {\cal A}^*}{ \partial \theta } \right]  , \nonumber 
\label{Generalized_LLE}
\end{eqnarray}
where $\eta_{\ell} = \partial \Lambda /\partial \ell$ for $\ell_m = \ell_n = \ell_p= \ell = \ell_0$, with $\eta_{\ell_m}$,  $\eta_{\ell_n}$, and  $\eta_{\ell_p}$ being defined analogously. 
The details of the derivation are analogous to similar derivations of the NLSE with higher-order corrections~\cite{Agrawal}. 
When the bandwidth of the resonator field approaches the carrier frequency, the Taylor expansion in Eq.~(\ref{frequency_expansion}) is no longer
valid, and one must use the full functional dependence of $\omega_\ell - \omega_0$ on $\ell$, by analogy to what is done in studies of supercontinuum generation.

In conclusion, we have demonstrated that the Kerr comb evolution in whispering gallery mode resonators
can be modeled using the Lugiato-Lefever equation and its extensions. We have shown that low-threshold,
wide-span combs can emerge as dissipative cavity solitons. With different parameters, rolls appear, corre-
sponding to narrower-band combs. So, when the goal is to optimize the bandwidth a correct parameter
choice is critical. Additionally, we demonstrated that the LLE can be extended to incorporate higher-order
dispersion and nonlinearity, and we expect that it can be extended without too much difficulty to include
Rayleigh, Brillouin, and Raman scattering~\cite{Boyd}. A key advantage of the spatio-temporal model that we have
developed is that the LLE has already been the subject of extensive mathematical study, and it should be
possible to take advantage of this earlier work to shed additional light on the dynamical properties of Kerr
combs. We expect that a better understanding of Kerr comb generation in whispering gallery resonators will
be the result, leading to new resonator designs that can produce octave-spanning combs.


Y. K. C acknowledges financial support from the European Research Council through
the project NextPhase (ERC StG 278616).  C. R. M acknowledges hospitality
and support from the FEMTO-ST Institute within the framework of the
LabEx Action.  C. R. M. also acknowledges useful discussions with T. Sylvestre.

\end{document}